\documentclass[twocolumn,prl,superscriptaddress,showpacs,byrevtex]{revtex4}
\usepackage{epsf,graphicx,amssymb}

\begin{document}
\title{Fluctuations of energy flux in wave turbulence}
\author{\'Eric Falcon}
\affiliation{Mati\`ere et Syst\`emes Complexes, Universit\'e Paris Diderot - Paris 7, CNRS, 75 013 Paris, France}

\author{S\'ebastien Auma\^\i tre}
\affiliation{Service de Physique de l'Etat Condens\'e, DSM, CEA-Saclay, CNRS, 91191 Gif-sur-Yvette, France}
\author{Claudio Falc\'on}
\affiliation{Laboratoire de Physique Statistique, \'Ecole Normale Sup\'erieure, CNRS, 24, rue Lhomond, 75 005 Paris, France}

\author{Claude Laroche}
\affiliation{Mati\`ere et Syst\`emes Complexes, Universit\'e Paris Diderot - Paris 7, CNRS, 75 013 Paris, France}
\affiliation{Laboratoire de Physique Statistique, \'Ecole Normale Sup\'erieure, CNRS, 24, rue Lhomond, 75 005 Paris, France}

\author{St\'ephan Fauve}
\affiliation{Laboratoire de Physique Statistique, \'Ecole Normale Sup\'erieure, CNRS, 24, rue Lhomond, 75 005 Paris, France}

\date{\today}

\begin{abstract}  
We report that the power driving gravity and capillary wave turbulence in a statistically stationary regime displays fluctuations much stronger than its mean value. We show that its probability density function (PDF) has a most probable value close to zero and involves two asymmetric roughly exponential tails. We understand the qualitative features of the PDF using a simple Langevin type model.  
\end{abstract}

\maketitle

When a dissipative system is driven in a statistically stationary regime by an external forcing, a given amount of power per unit mass, $\epsilon$, is transfered from the driving device to the system and is ultimately dissipated. In fully developed turbulence, a flow is driven at large scales and nonlinear interactions transfer kinetic energy toward small scales where viscous dissipation takes place. In the intermediate range of scales (the inertial range) the key role of the energy flux $\epsilon$ has been first understood by Kolmogorov \cite{Kolmogorov41}. Using dimensional arguments, he derived the law $E(k) \propto \epsilon^{2/3} k^{-5/3}$ for the energy density $E(k)$ as a function of the wavenumber $k$. Kolmogorov type spectra have been derived analytically in wave turbulence, i.e. in various systems involving an ensemble of weakly interacting nonlinear waves (see for instance \cite{turbulonde} for a review). In all cases, it has been assumed that $\epsilon$ is a given constant parameter. However, it should be kept in mind that $\epsilon$ is not an input parameter in most experiments or simulations of dissipative systems. Its value is not externally controlled but determined by the impedance of the system. In addition, as we have already shown for a variety of different dissipative systems \cite{global,Aumaitre01,Aumaitre04}, the energy flux or related global quantities, strongly fluctuate in time although being averaged in space on the whole system or on its boundaries. These fluctuations should not be confused with small scale intermittency which occurs in fully developed turbulence. The later is related to the spotness of dissipation in space \cite{Kolmogorov62} and its description does not involve a time dependent $\epsilon$.

Here we study the fluctuations of the injected power in wave turbulence. Gravity-capillary waves are generated on a fluid layer by low frequency random vibrations of a wave maker. By measuring the applied force on the wave maker and its velocity, we determine the instantaneous power $I(t)$ injected into the fluid. We observe that it strongly fluctuates. Its most probable value is $0$. $rms$ fluctuations $\sigma_I$ up to several times the mean value $\langle I \rangle$ are observed, and the probability density function (PDF) of $I$ displays roughly exponential tails for both positive and negative values of $I$. These negative values correspond to events for which the random wave field gives back energy to the driving device. We show how fluctuations of the injected power depend on the system size and on the mean dissipation and we study their statistical properties. 

\begin{figure}[!htb]
\centerline{
\epsfysize=60mm
\epsffile{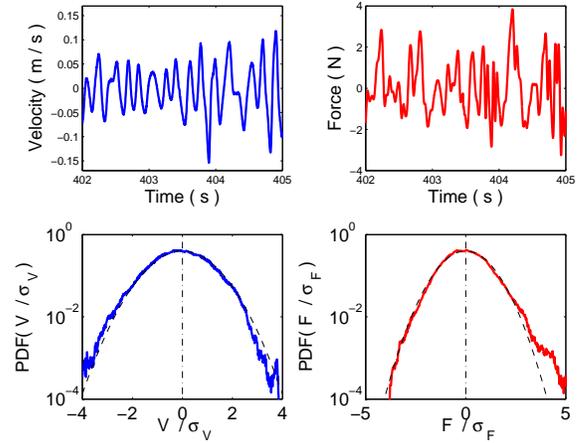}
}
\caption{(color online) Time recordings of the velocity of the wave maker and the force applied to the wave maker by the vibration exciter. The fluid is mercury, with  $h=23$ mm. Both PDF are Gaussian (dashed lines) with zero mean value.}
\label{fig01}
\end{figure}

The experimental setup, described in \cite{Falcon07}, consists of a rectangular plastic vessel, with lateral dimensions $57 \times 50$ or $20 \times 20$ cm$^2$, filled with water or mercury (density $13.6$ times larger than water) up to a height, $h = 1.8$ or $2.3$ cm. Surface waves are generated by the horizontal motion of a rectangular ($L\times H$ cm$^2$) plunging PMMA wave maker driven by an electromagnetic vibration exciter.  We take $1.2 < L < 25$ cm and $H=3.5$ cm. The wave maker is driven with random noise excitation below $4$ or $6$ Hz. 

The power injected into the wave field by the wave maker is determined as follows. The velocity $V(t)$ of the wave maker is measured using a coil placed on the top of the vibration exciter. The voltage induced by the moving permanent magnet of the vibration exciter is proportional to $V(t)$. The force $F_A(t)$ applied by the vibration exciter on the wave maker is measured by a piezoresistive force transducer (FGP 10 daN). 
The time recordings of  $V(t)$ and $F_A(t)$ together with their PDFs are displayed in Fig. 1. 
Both $V(t)$ and $F_A(t)$ are Gaussian with zero mean value. For a given excitation bandwidth, the $rms$ value $\sigma_V$ of the velocity fluctuations of the wave maker is proportional to the driving voltage $U$ applied to the electromagnetic shaker and does not depend on the fluid density $\rho$. On the contrary, the standard deviation $\sigma_{F_A}$ of the force applied to the wave maker is decreased by the density ratio ($\sim 13$) when mercury is replaced by water. 
We have checked that $\sigma_{F_A} \propto \rho S_P \sigma_V$ where $S_P=Lh$ is the immersed area of the wave maker. This linear behavior has been measured on one decade up to $\sigma_{F_A} \sim 2$ N and $\sigma_V \sim 0.1$ m/s.

\begin{figure}[!htb]
\centerline{
\epsfysize=65mm
\epsffile{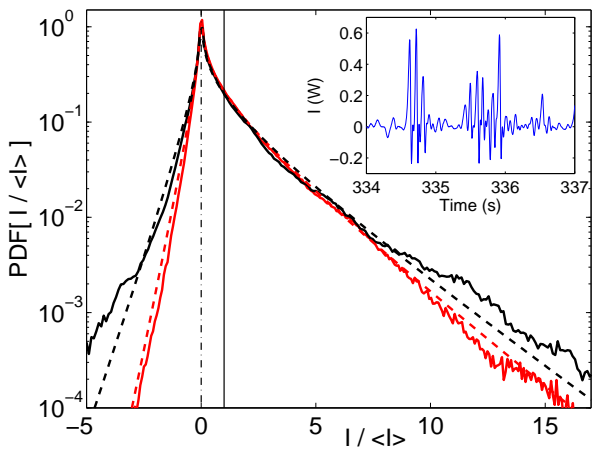}
}
\caption{(color online) PDF of $I(t)/\langle I \rangle$ for mercury: container size $57 \times 50$ cm$^2$ (red) and $20 \times 20$ cm$^2$ (black) ($h=18$ mm). Dashed lines are the related predictions from equation (\ref{PDFI}) without fitting parameter. Inset: time recording of $I(t)$.}
\label{fig02}
\end{figure}

When the wave maker inertia is negligible, the power $I(t)$ injected into the fluid is roughly given by $F_A(t) V(t)$ (see below). The time recording of $I(t)$  is shown in the inset of Fig. 2. Contrary to the velocity or the force, the injected power consists of strong intermittent bursts. Although the forcing is statistically stationary, there are quiescent periods with a small amount of injected power interrupted by bursts where $I(t)$ can take both positive and negative values. The PDFs  of $I/ \langle I \rangle$ are displayed in Fig. 2.  They show that the most probable value of $I$ is zero and display two asymmetric exponential tails (or stretched exponential in the smaller container). We observe that events with $I(t) < 0$, {\it i.e.} for which the wave field gives back energy to the wave maker, occur with a fairly high probability.
The standard deviation $\sigma_I$ of the injected power is much larger than its mean value $\langle I \rangle$ and rare events with amplitude up to $7 \, \sigma$ are also detected. Typical values obtained when $\sigma_V \sim 0.05$ m/s are $\sigma_{F_A} \sim 1$ N, $\langle I \rangle \sim 30$ mW, $\sigma_I \sim 100$ mW for mercury. Our measurements also show that $\sigma_I \propto \langle I \rangle = c \rho S_P \sigma_V^2$, 
where $c$ has the dimension of a velocity ($c \sim 0.5$ m/s and slightly increases when the container size is increased). 

We also observe in Fig. 2 that the probability of negative events strongly decreases when the container size is increased whereas the positive fluctuations are less affected. This shows that the backscattering of the energy flux from the wave field to the driving device is related to the waves reflected by the boundary that can,  from time to time, drive the wave maker in phase with its motion.  We note that we have less statistics for the negative tail of the PDF when the size of the container is increased. 

We recall that the statistical properties of the fluctuations of the surface height have been studied in \cite{Falcon07}: they involve a large distribution of amplitude fluctuations. Their frequency spectrum is broad band and can be fitted by two power laws in the gravity and capillary regimes. The power law exponent in the capillary range is in agreement with theoretical predictions. The one in the gravity range depends on the forcing, as also shown in \cite{Nazarenko07}. The scaling of the spectrum with respect to the mean energy flux $\langle I \rangle$ is different from the theoretical prediction both in the gravity and capillary ranges. These discrepancies can be ascribed to finite size effects \cite{Falcon07,Nazarenko07}.

\begin{figure}[!htb]
\centerline{
\begin{tabular}{cc}
\epsfxsize=45mm \epsffile{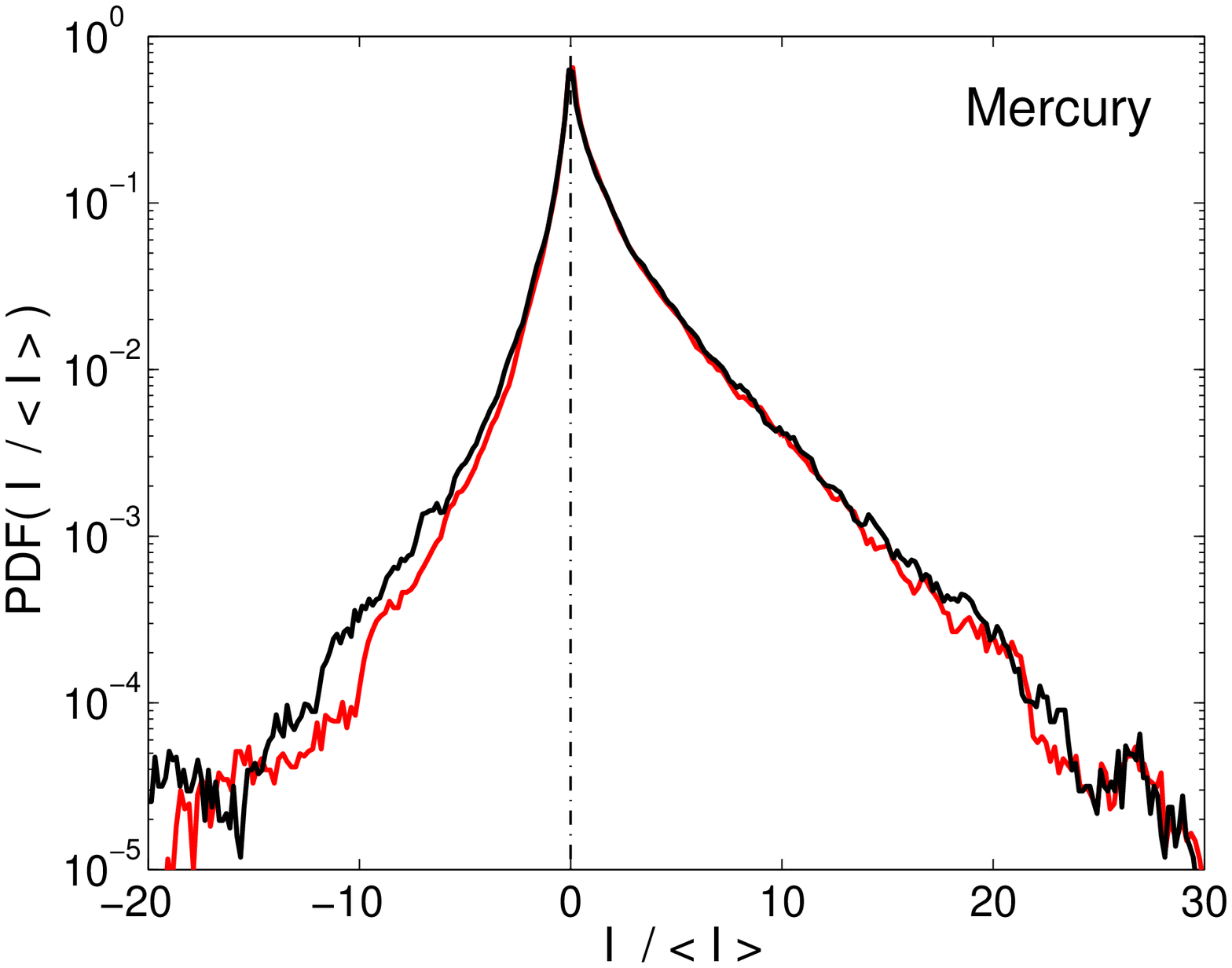} 
& 
\epsfxsize=45mm \epsffile{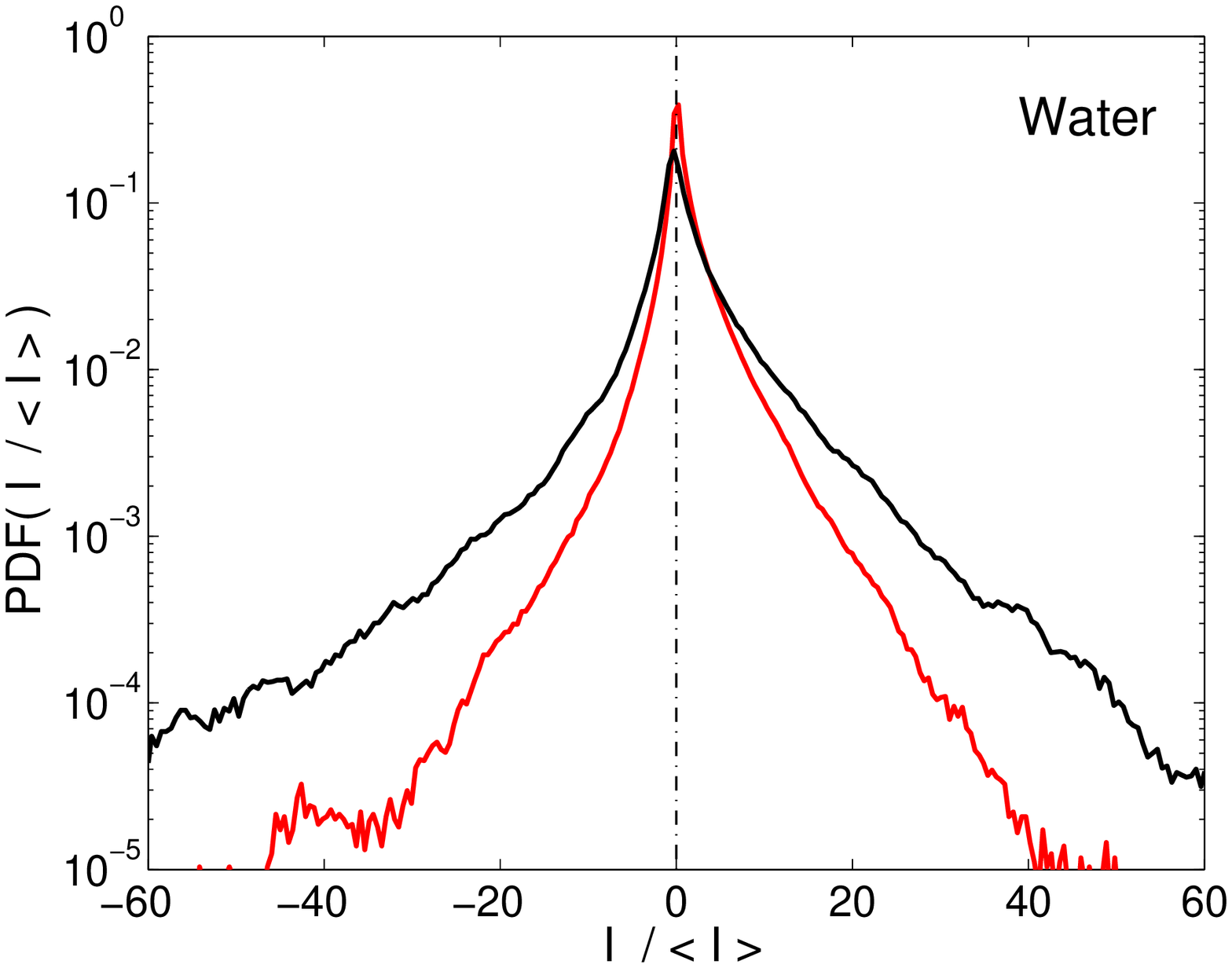} 
\end{tabular}}
\caption{(color online) Effect of the inertia of the wave maker: PDF of $F_A V$ (black) and of $(F_A -M\, \dot V)V$ (red) for mercury ($h=23$ mm) (left) and for water ($h=23$ mm) (right). Using $F_A V$ to estimate $I$ leads to an error on the standard deviation $\sigma_I$ that is less than $5 \%$ for mercury but that reaches $50 \%$ for water.}
\label{fig03}
\end{figure}

We first emphasize the bias that can result from the system inertia when one tries a direct measurement of the fluctuations of injected power.
The equation of motion of the wave maker is 
\begin{equation}
M \dot V = F_A(t) + F_R (t),
\label{wavemaker}
\end{equation}
where $M$ is the mass of the wave maker and $F_R (t)$ is the force due to the fluid motion ($\dot V \equiv dV/dt$). The power injected into the fluid by the wave maker is $I(t) = -F_R (t) V(t)$. When $M\, \dot V$ is not negligible, $I(t)$ generally differs from $F_A(t) V(t)$ which is experimentally determined. This obviously does not affect the mean value $\langle I \rangle$ but may lead to wrong estimates of fluctuations. Using an accelerometer, we have checked that $M\, \dot V$ is negligible compared to $F_A$ when the working fluid is mercury. This is shown is Fig. 3 (left) where the PDF of $F_AV$ and $F_R V = (F_A-M\, \dot V)V$ are compared.  On the contrary, inertia cannot be neglected for experiments in water for which an error as large as one order of magnitude can be made on the probability of rare events if one use $F_A V$ to estimate $I$ (right). Thus, the correction due to $M\, \dot V$ has been taken into account in water experiments.
There exist only a few previous direct measurements of injected power in turbulent flows and those type of inertial bias have never been taken into account \cite{vkbias}.

\begin{figure}[!htb]
\centerline{
\epsfysize=50mm
\epsffile{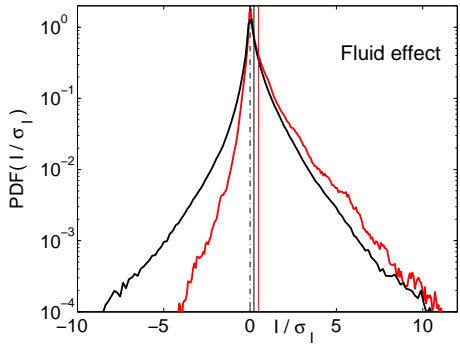}
}
\caption{(color online) Effect of fluid properties on the PDF of the injected power: (red) mercury; (black): water ($h = 18$ mm; 20$\times$20 cm$^2$ container). Solid lines indicate the value of $\langle I \rangle / \sigma_I$.}
\label{fig04}
\end{figure}

The PDFs of injected power for the same driving in the same container for water and mercury are displayed in Fig. 4. The asymmetry of the PDF is much larger with mercury. This is related to its larger mean energy flux, i.e., mean dissipation, as shown below.

The qualitative features of the PDF of injected power can be described with the following simple model.  Guided by our experimental observation of the linearity of $\sigma_{F_A}$ in $\sigma_V$, we assume that the force $F_R$ due to the fluid can be roughly approximated by a friction force $-M \gamma V$ where $\gamma$ is a constant (the inverse of the damping time of the wave maker). We are aware that a better approximation to the force due to the fluid should involve both $\dot V$ and an integral of $V(t')$ with an appropriate kernel. Thus we only claim here to give a heuristic understanding of the qualitative properties of the PDF of $I$. Modelling the forcing with an Ornstein-Uhlenbeck process, we obtain

\begin{equation}
\dot V = - \gamma V + F, \;\;\;
\dot F = - \beta F + \xi,
\label{OU}
\end{equation}
where $\beta$ is the inverse of the correlation time of the applied force ($F=F_A/M$) and $\xi$ is a Gaussian white noise with $\langle \xi(t) \xi(t') \rangle = D 
\delta (t-t')$. The PDF $P(V, F)$ is the bivariate normal distribution \cite{Risken,Bandi07}
\begin{equation}
P(V, F)=
\frac{\exp {\left[- \frac{1}{2 (1-r^2)} \left(\frac{V^2}{\sigma_V^2} - \frac{2 r VF}{\sigma_V \sigma_F} + \frac{F^2}{\sigma_F^2}\right)\right]}     }{2\pi \sigma_V \sigma_F \sqrt{1-r^2}},
\label{bivariate}
\end{equation}
with $\sigma_F = \sqrt{D/2\beta}$, $\sigma_V = \sqrt{D/(2\gamma \beta (\gamma+ \beta))}$ and $r = \sqrt{\gamma/(\gamma + \beta)}$. Changing variables $(V, F)$ to $(\tilde I = FV = I/M, F)$ and integrating over $F$ gives
\begin{equation}
P(\tilde I) = \frac{\exp \left[\frac{r \tilde I}{(1-r^2) \sigma_V \sigma_F}\right]}{\pi \sigma_V \sigma_F \sqrt{1-r^2}}\,  \, K_0 \left[\frac{\vert \tilde I \vert}{(1-r^2) \sigma_V \sigma_F}\right],
\label{PDFI}
\end{equation}
 where $K_0 (X)$ is the zeroth order modified Bessel
function of the second kind. Using the method of steepest descent, this predicts exponential tails, $P(X) \sim (1/\sqrt{\vert X \vert}) \exp (rX - \vert X \vert)$ where $X = \tilde I / [(1-r^2) \sigma_V \sigma_F]$. In addition, we have $\langle \tilde I \rangle = D/[2\beta(\gamma + \beta)] = r \sigma_V \sigma_F$. Thus, (\ref{PDFI}) is determined once $\langle I \rangle$, $\sigma_V$ and $\sigma_F$ have been measured and can be compared to the experimental PDF without using any fitting parameter. This is displayed with dashed lines in Fig. 2. Taking into account the strong approximation made in the above model, we observe a good agreement in the larger container. More importantly, this model captures the qualitative features of the PDF: its maximum for $I = 0$ and the asymmetry of the tails that is governed by the parameter $r = \sqrt{\gamma/(\gamma + \beta)} = \langle I \rangle /(\sigma_V \sigma_{F_A})$.  For given $\sigma_V$ and  $\sigma_{F_A}$, the larger is the mean energy flux, i.e., the dissipation, the more asymmetric is the PDF.  For mercury, direct determination of $r$ from the measurement of  $\langle I \rangle$, $\sigma_V$ and $\sigma_{F_A}$ gives $r \sim 0.7$ for the large container and $r \sim 0.6$ for the small one, in qualitative agreement with the different asymmetry of the PDF in Fig. 2. Smaller values of $r$ are achieved in water for which the dissipation is smaller. The PDFs are more stretched for water in particular in the smaller container. 
  
\begin{figure}[!htb]
\centerline{
\epsfysize=70mm
\epsffile{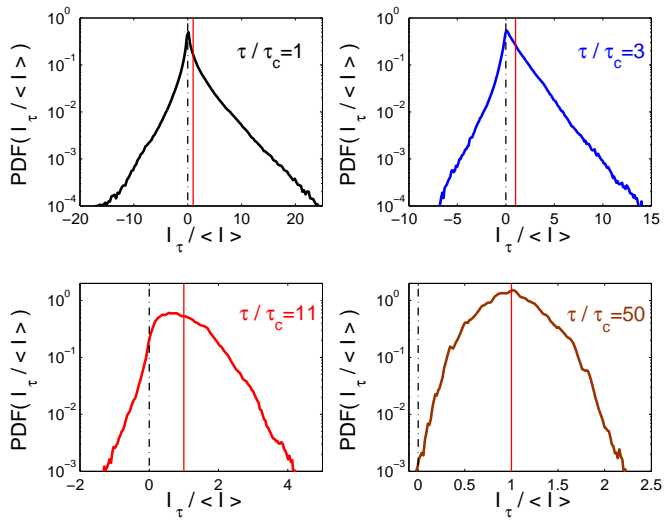} 
}
\caption{(color online) PDFs of the injected power $I_{\tau}$ averaged on a time interval $\tau$:  $\tau = 1, 3, 11$ and $50 \tau_c$, where $\tau_c = 0.03$ s is the correlation time of $I(t)$. Solid lines indicate the value of $\langle I \rangle$ (water, $h=23$ mm).}
\label{fig5}
\end{figure}

We now consider the injected power averaged on a time interval $\tau$
\begin{equation}
I_{\tau} (t) = \frac{1}{\tau} \int_t^{t+\tau} I(t') dt'.
\end{equation}
The PDFs of $I_{\tau}$ for $\tau/\tau_c = 1, 3, 11$ and $50$ where $\tau_c$ is the correlation time of $I(t)$, are displayed in Fig. 5. They become more and more peaked around $I_{\tau} \simeq \langle I \rangle$ as they should. However, one needs to average on a rather large time interval ($\tau \sim 50 \tau_c$) in order to get a maximum probability $P(I_{\tau})$ for $I_{\tau} = \langle I \rangle$ (Fig. 5, bottom right). Then, the probability of negative events become so small that almost none can be observed.  
Fig. 6 shows that the quantity, $\frac{1}{\tau}\,\log {\frac{P(I_{\tau}/\langle I \rangle)} {P(- I_{\tau}/ \langle I \rangle)}}$ for different values of $\tau$ that has been predicted to be linear in $I_{\tau}/ \langle I \rangle$ when the hypothesis of the fluctuation theorem (in particular time reversibility) are fulfilled \cite{FT,Kurchan}.  As we clearly observe in Fig. 6, this is not  the case in general for dissipative systems.  As already mentioned \cite{Aumaitre01} and studied in details \cite{orsayboys}, the linear behavior reported in several experiments or numerical simulations results from the too small values of $I_{\tau}/ \langle I \rangle$ that are probed when $\tau \gg \tau_c$. Large enough values are obtained in the present experiment and the expected nonlinear behavior is thus reached. The shape of the curve in Fig. 6 is found in good agreement with the analytical calculation \cite{Farago} performed with a Langevin type equation with white noise. 
 
\begin{figure}[!htb]
\centerline{
\epsfysize=60mm
\epsffile{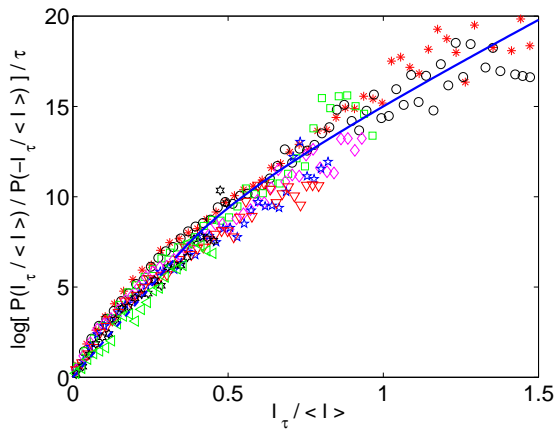}
}
\caption{(color online) Plot of $\frac{1}{\tau}\,\log {\frac{P(I_{\tau}/ \langle I \rangle)} {P(- I_{\tau} / \langle I \rangle)}}$ for $16 < \tau/\tau_c < 39$ [$\tau /\tau_c = 17$ ($\ast$),   19.5 ($\circ$),  22 ($\square$),  25 ($\lozenge$), 28 (pentagram), 30.5 ($\triangledown$), 33.5 (hexagram), 39 ($\triangleleft$)]. Langevin model of reference \cite{Farago}: $4\gamma$ for $\epsilon = I_\tau / \langle I \rangle \leq 1/3$ (dashed line) and $7\gamma\epsilon/4 +3\gamma/2-\gamma/(4\epsilon)$ for $\epsilon \geq 1/3$ (solid line) with $\gamma=5$ Hz.}
\label{fig06}
\end{figure}

Finally, we emphasize that a fluctuating injected power implies fluctuations of the energy flux at all wave numbers in the energy cascade from injection to dissipation. 
In any system where an energy flux cascades from the injected power at large scales to dissipation at small scales, one has for the energy $E_{<}$ for wave numbers smaller than $k$ within the inertial range, $\dot E_{<} = I(t) - \Phi (k,t) \equiv R$,
where $\Phi(k, t)$ is the energy flux at $k$ toward large wave numbers.
Thus $\int_0^\infty \langle R(\tau) R(0) \rangle d\tau =0$ in order to prevent the divergence of $\langle E_{<}^2 \rangle$. Dimensionaly, this implies that $\sigma^2_{\Phi} \tau_k$ does not depend on $k$ \cite{Aumaitre04}, where $\sigma_{\Phi}$ is the standard deviation of the energy flux and $\tau_k$ is its correlation time. If this dimensional scaling is correct, fluctuations of the energy flux are expected to increase during the cascade from large to small scales since $\tau_k$ decreases (for instance, $\sigma_{\Phi} \propto k^{1/3}$ for hydrodynamic turbulence). 
Such fluctuations have been found numerically and experimentally in hydrodynamic turbulence \cite{Eflux}. To which extent, this is related or modified by small scale intermittency \cite{Falcon07b} remains an open question.

We acknowledge useful discussions with F. P\'etr\'elis. This work has been supported by ANR turbonde BLAN07-3-197846 and by the CNES.\\


\end{document}